\documentclass{amsart}

\newtheorem{thm}{Theorem}[section]

 \newtheorem{prop}[thm]{Proposition}
 \theoremstyle{definition}
 \newtheorem{defn}[thm]{Definition}
 \theoremstyle{remark}
 
\numberwithin{equation}{section}

 \newcommand{\D}{\mathfrak{D}}
 \newcommand{\dd}{\rm d}
 
 \newcommand{\F}{\mathfrak{F}}
 \newcommand{\h}{\mathfrak{H}}
 
 \newcommand{\T}{\mathfrak{T}}
 
 \newcommand{\DPRM}{\textsc{dprm }}

\usepackage{amsfonts}
\usepackage{graphicx}
\usepackage{hyperref}

\begin{document}


\title[Hypercomputability of Quantum
Adiabatic Processes]{Hypercomputability of Quantum Adiabatic
Processes:\\Facts versus Prejudices}


\author{Tien D. Kieu}
\email[Tien D. Kieu]{kieu@swin.edu.au}
\address{Centre for Atom Optics and Ultrafast Spectroscopy,
Swinburne University of Technology, Hawthorn 3122, Australia}


\begin{abstract}
We give an overview of a quantum adiabatic algorithm for Hilbert's
tenth problem, including some discussions on its fundamental aspects
and the emphasis on the probabilistic correctness of its findings.
For the purpose of illustration, the numerical simulation results of
some simple Diophantine equations are presented. We also discuss
some prejudicial misunderstandings as well as some plausible
difficulties faced by the algorithm in its physical implementations.
\end{abstract}

\maketitle

\begin{flushright}
{\em ``To believe otherwise is merely to cling to a prejudice\\which
only gives rise to further prejudices $\ldots$"}~\footnote{An ending
statement for the book {\em Twenty Lectures on Thermodynamics},
(Pergamon Press, Oxford, 1975), warning
against the prejudice that negative absolute temperatures could not be defined.}\\
H.A. Buchdahl
\end{flushright}

\section{Introduction and outline of the paper}
Quantum computation has attracted much attention and investment
lately through its theoretical potential to speed up some important
computations as compared to classical Turing computation. The most
well-known and widely-studied form of quantum computation is the
(standard) model of quantum circuits~\cite{qcbook} which comprise
several unitary quantum gates (belonging to a set of universal
gates), arranged in a prescribed sequence to implement a quantum
algorithm. Each quantum gate only needs to act singly on one qubit,
the generalisation of the classical bit, or on a pair of qubits.

In view of this promising potential of quantum computation, another
question that then inevitably arises is whether the class of Turing
computable functions can be extended by applying quantum principles.
This kind of computation beyond the Turing barrier is also known as
hypercomputation~\cite{Copeland, Toby02, Toby05}. Initial efforts
have indicated that quantum computability is the same as classical
computability~\cite{bernsteinetal}. However, this negative
conclusion is only valid for the standard quantum computation. And
we know that such a standard model is not the only model of quantum
computation. Nor does it necessarily exploit fully the principles of
quantum mechanics.

As an illustration of the ability of quantum mechanics, a truly
random sequence of bits could be easily generated from a qubit
initially prepared in a certain state, while Turing machines have to
be content with only pseudo-random generators (see, for
example,~\cite{Stannett}). (Thus, it seems from the view point of
Algorithmic Information Theory~\cite{Chaitin:1987} that a finitely
prepared qubit can have an infinite algorithmic informational
theoretic complexity as compared to any finite Turing machine!) This
ability to generate random numbers could thus be defined as a form
of hypercomputation -- albeit of very limited application and of a
nature which is non-harnessible or non-exploitable for universal
computation.

Also, not restricted by the standard model of quantum computation,
we have claimed~\cite{kieu-contphys, kieu-spie, kieuFull,
kieu-intjtheo, kieu-royal} to have a quantum algorithm for Hilbert's
tenth problem~\cite{hilbert10} despite the fact that the problem has
been proved to be recursively noncomputable. The algorithm makes
essential use of the Quantum Adiabatic Theorem (QAT)~\cite{messiah}
and other results in the framework of Quantum Adiabatic Computation
(QAC)~\cite{qac}, subject to some predetermined and arbitrarily
small probability of inaccuracy. We name this kind of algorithms
{\em Probabilistically Correct Algorithms} to emphasise the fact
that the end results from such algorithms are subject to some {\em
probabilities of being incorrect}. Such error probabilities are
necessary when there is, in principle, no other way to verify all
the outcomes of the algorithms.

QAC represents another model of quantum computation, alternative to
the standard model, and recent studies indicate that QAC offers
certain advantages in error and decoherence control~\cite{farhi2}
and in experimental realisation via quantum optical
implementations~\cite{Lloyd}. More interestingly, the application of
QAC to infinite-dimension or unbounded spaces has the potential to
extend the classical notion of computability, that is, potential for
hypercomputation, through a positive resolution of the Turing
halting, Hilbert's tenth and associated recursively noncomputable
problems. For some 70 years many have speculated on some possible
extensions of the notion of Church-Turing
computability~\cite{Copeland}. The recognition of probabilistic
quantum physics as the missing ingredients may now be the key for
such achievement.

In the next Section we introduce Hilbert's tenth problem and its
equivalence to the Turing halting problem together with their
finitely refutable, but Turing noncomputable, characters. Then we
collect in the following Section the key ingredients of our proposed
quantum adiabatic algorithm for Hilbert's tenth problem. Emphasis
will be put on the probabilistically correct nature of the algorithm
and how this nature could evade the no-go results stemmed from
Cantor's diagonal arguments (see~\cite{OrdKieu-diag}, for example).

For illustration purposes, we present in Section~\ref{numerical}
some numerical simulation examples employing the algorithm with
speculation how the infiniteness of the underlying Hilbert space
could be probabilistically handled on finite Turing machines. We
then move on to a discussion of the misunderstandings or prejudices
that we know of, and some potential problems in the physical
implementation of the algorithm. The paper concludes with some
remarks.

\section{Hilbert's tenth problem and its significance}
\subsection{Hilbert's tenth problem}
In 1900, David Hilbert presented a list of important problems for
the mathematicians of the coming twentieth century. Of twenty three
problems on the list, there was only one decision problem
(entcheidungsproblem) and also the only one that was not given a
name. Ever since it has been called Hilbert's tenth
problem~\cite{hilbert10}, being its position on the list,
\begin{defn}{(Hilbert's tenth problem)}
\em Given any polynomial equation with any number of unknowns and
with integer coefficients: To devise a universal process according
to which it can be determined by a finite number of operations
whether the equation has (non-negative) integer solutions.
\label{tenth}
\end{defn}
The polynomial equations in question are also called Diophantine
equations, some examples of which are the ones involved in Fermat's
last theorem, the most celebrated and well known problem of number
theory. Clearly, the availability of such a procedure in the problem
of Definition~(\ref{tenth}) is more than a convenience; it would
also simplify mathematics and at the same time remove some of the
creativeness required of mathematicians for different Diophantine
equations. Indeed, if the universal procedure Hilbert asked for is
available, many important mathematical problems and conjectures
would be exposed and settled by a {\em single mechanical procedure}.
These are the problems and conjectures which have some Diophantine
representations and whose truth values depend on the existence or
the lack of (non-negative) integer solutions of the characteristic
Diophantine equations. Important examples of these
include~\cite{mountaintop}:
\begin{itemize}
\item The Four-Colour Problem: Whether only four colours are
sufficient to colour a world map in such a way that no two adjacent
countries share the same colour.
\item The Goldbach conjecture: Any positive even number can be
written as the sum of two primes.
\item The Riemann Hypothesis involving the distribution of
(non-trivial) zeros of the Riemann zeta function. Determination the
truth value of this conjecture is considered by some to be the most
important problem of contemporary mathematics.
\end{itemize}
Associated with each of the above example is a (complicated)
Diophantine equation, from which the determination of the truth
value for the conjecture is reduced simply to the determination of
whether the corresponding equation has a solution or not -- a task
which could be made much simpler mechanically with Hilbert's wish.

The significance of Hilbert's tenth problem is not only limited to
the above. It is intimately linked to the formalisation of
mathematics (a dream harboured by Leibnitz~\cite{Davis02,
ChaitinBook05} in the seventeenth century and later by Hilbert
himself), a dream which was shattered by G\"odel's incompleteness
theorem and, later, the Turing halting theorem. When proposing the
tenth problem in 1900, Hilbert himself never anticipated the link it
would have with what is the Turing halting problem of the
yet-to-be-born field of Theoretical Computer Science. The Turing
halting problem was only introduced and solved in 1937 by Turing,
and the equivalence of the two problems was not established until
1972~\cite{hilbert10}, culminating the efforts of three generations
of mathematicians.

\subsection{\label{no-go}Turing halting theorem}
The question of the Turing halting problem concerns whether there
exists a universal process according to which it can be determined
by a finite number of operations if any given Turing machine would
eventually halt (in finite time) starting with some specific input.
Turing raised this problem in parallel similarity to the G\"odel's
Incompleteness Theorem and showed that there exists no such
recursive universal procedure. His proof is based on Cantor's
diagonal arguments, also employed in the proof of the Incompleteness
Theorem.

The proof is by contradiction, starting with the assumption that
there exists a Turing computable halting function $h(p,i)$ which
accepts two integer inputs: $p$, the G\"odel encoded integer for the
Turing machine in consideration, and $i$, the G\"odel encoded
integer for the input for $p$,
\begin{eqnarray}
h(p,i)&=&\left\{
\begin{tabular}{l}
0 \mbox{ if $p$ halts on input $i$};\\
1 \mbox{ otherwise}.
\end{tabular}
\right. \label{Cantor}
\end{eqnarray}
One can then construct a program $Turing(n)$ having one integer
argument $n$ in such a way that it calls the function $h(n,n)$ as a
subroutine and then halts if and only if $h(n,n) = 1$.  In some
made-up language:
\begin{quotation}
\begin{tabular}{ll}
 &\tt Program $Turing$\\
 &\tt input $n$\\
10& \tt call $h(n,n)$\\
 &\tt if $h(n,n) = 0$ goto 10\\
 &\tt stop\\
 &\tt end
\end{tabular}
\end{quotation}

Let $t$ be the G\"odel encoded integer for the program $Turing$; we
now apply the assumed halting function $h$ to $t$ and $n$, then
clearly:
\begin{eqnarray}
h(t,n) = 0 && \mbox{\bf  iff }
Turing \mbox{ halts on } n\nonumber\\
&&\mbox{\bf iff } h(n,n) =1. \label{T1}
\end{eqnarray}
A contradiction is clearly manifest once we choose $n = t$ in the
above. Thus there cannot exist any Turing machine that could compute
the halting function $h(p,i)$:
\begin{thm}[The Halting Theorem]
For any Turing machine program $h$ purporting to settle the halting
or non-halting of all Turing machine programs, there exists a
program $p$ and input data $i$ such that the program $h$ cannot
determine whether $p$ will halt when processing the data $i$.
\end{thm}
The above theorem is stated in a form~\cite{CastiDePauli} that
highlights its similarity to G\"odel's theorem:
\begin{thm}[G\"odel's Incompleteness Theorem]
For any consistent formal system $\F$ purporting to settle -- that
is, prove or disprove -- all statements of arithmetic, there exists
an arithmetical proposition that can be neither proved nor disproved
in this system.  Therefore, the formal system $\F$ is incomplete.
\end{thm}

\subsection{The equivalence of the two problems}
It is easy to see that if one can solve the Turing halting problem
then one can solve Hilbert's tenth problem. This is accomplished by
constructing a simple program that systematically searches for the
zeros of a given Diophantine equation by going through the tuples of
non-negative integers one by one and stops only when a solution is
found. The Turing halting function (existed by assumption) can then
be applied to that particular program to find out if it ever halts.
It halts if and only if the Diophantine equation has a non-negative
integer solution.

Proving the relationship in the opposite direction, namely that if
Hilbert's tenth problem can be solved then will be the Turing
halting problem, is much more difficult and requires the so-called
Davis-Putnam-Robinson-Matiyasevich (\DPRM) Theorem~\cite{hilbert10}:
\begin{thm}[\DPRM Theorem]
Every recursively enumerable (r.e.)~set $\D$~\footnote{A set $\D$ is
recursively enumerable if it is the range of an unary recursive
function $d: {\mathbb N} \to {\D}$. In other words, a set is
recursively enumerable if there exists a Turing program which
semi-computes it -- that is, when provided with an input, the
program returns 1 if that input is an element of the set, and
diverges otherwise.} of $n$-tuples of non-negative integers has a
Diophantine representation, that is:
\begin{equation}
\langle a_{1},\cdots,a_{n} \rangle  \in \D \quad \Longleftrightarrow
\quad \exists\,
x_{1}\cdots{}x_{m}:\,{D(a_{1},\cdots,a_{n};\,x_{1},\cdots,x_{m})=0}
\end{equation}
\end{thm}
As an interesting diversion, we note that when a set is not r.e.~the
\DPRM Theorem is not directly applicable; but in some special cases
the Theorem can still be very useful. One such interesting example
is the set whose elements are the positions (in a given ordering) of
all the bits of Chaitin's $\Omega$~\cite{Chaitin:1987} which have
value 0 (in some fixed programming language). We refer the reader
to~\cite{ordkieu-omega, ChaitinBook05} for further exploitation of
the \DPRM Theorem in representing the bits of $\Omega$ by some
properties (be it the parity or the finitude) of the number of
solutions of some Diophantine equations. This representation for
$\Omega$ has recently been extended further by
Matiyasevich~\cite{Matiyasevich04}.

Back to the equivalence between the Turing and Hilbert's tenth
problems, let us number all Turing machines (that is, programs in
some fixed programming language) uniquely in some lexicographical
order, say. The set of all non-negative integer numbers
corresponding to all Turing machines that will halt when started
from the blank tape is clearly a r.e.~set. Let us call this set the
halting set, and thanks to the \DPRM Theorem above we know that
corresponding to this set there is a family of one-parameter
Diophantine equations. If Hilbert's tenth problem were recursively
soluble, that is, were there a recursive method to decide if any
given Diophantine equation has any solution then we could have
recursively decided if any Turing machine would halt when started
from the blank tape. We just need to find the number representing
that Turing machine and then decide if the relevant Diophantine
equation having the parameter corresponding to this number has any
solution or not. It has a solution if and only if the Turing machine
halts.

But that would have contradicted the Cantor's diagonal arguments for
the recursive noncomputability of Turing halting problem!  Thus, one
comes to the conclusion that there is no single recursive method for
deciding Hilbert's tenth problem.   To determine the existence or
lack of solutions of each different Diophantine equation one would
need different (recursive) methods anew each time.

The elegant proof above was only intended by Turing for the
non-existence of a Turing computable halting function.
Unfortunately, some have used these arguments to rule out any kind
of hypercomputation beyond the capability of Turing
machines~\cite{Cotogno}! We have considered carefully the implicit
assumptions of Cantor's diagonal arguments and pointed out
elsewhere~\cite{OrdKieu-diag} (see also~\cite{Welch04}) the
fallacies of such misuse of Cantor's diagonlisation. In particular,
we wish to point out, and will substantiate in the below, that
logically there is nothing wrong with non-recursive or
non-deterministic or probabilistic methods for deciding Hilbert's
tenth problem.

\subsection{\label{finitely}Finitely refutable but non-computable problems}
Hilbert's tenth problem is actually semi-decidable in the sense that
a systematic substitution of the non-negative ($n$-tuple) integers
into a Diophantine equation would eventually yield a solution, if
that equation indeed has a solution. But if the equation has no
solution at all, one might think that such substitution process
would go on {\em ad infinitum} and could never terminate.

As a matter of fact, for each given Diophantine equation, if it has
no solution in a {\em finite} domain then it has no solution at all
in the infinite domain of non-negative integers! Such problems are
thus termed finitely refutable~\cite{CaludeBookRandom}.
Nevertheless, Hilbert's tenth problem is still recursively
non-computable simply because there exists no universal recursive
procedure to determine the different finite domain for each and
every Diophantine equation.

The busy beaver function provides an example of the existence of
such finite but undecidable upper bounds for Turing computability as
the function grows faster than any Turing computable function. More
precisely, a decision problem can be posed as the question whether
an integer $j\in \D$ or not, where $\D$ is a standard recursively
enumerable non-recursive set. Because it is recursively enumerable
there exists a total computable function $d$ from the set of natural
numbers to itself, $d: {\mathbb N} \to {\mathbb N}$, which
enumerates $\D$ without repetitions. A special function called the
{\em waiting-time} function for $d$ is defined as
\begin{eqnarray}
\nu(j) &=& \mu\, n\, [d(n) = j\,]; \label{waiting}
\end{eqnarray}
that is, $\nu(j)$ gives the least $n$ (`time') upon which $j$ is
confirmed to be a member of the set $\D$. Note that $\nu$ is a
partial recursive function which is not bounded by any total
computable function. If one knows a bound $B$ for the total function
\begin{eqnarray}
\beta(J) &=& \max \,\{\nu(j): j<J \;\&\; j\in \D\},
\end{eqnarray}
which is the maximum waiting time for $d(n)$ within the range $J$,
then one can always give correct answers to the question whether
$j\in \D$, for all $j<J$. Indeed, if the bound $B$ is known one just
need to compute $d(n)$ only for all $n<B$ to see if $j$ (where
$j<J$) is obtained as a value of $d$, and thus belongs to $\D$ or
not. However, the non-computability of our problem lies in the fact
that $\beta(J)$, even though total, is noncomputable: it eventually
majorises every recursively computable function~\footnote{Gandy has
applied this function to challenge the hypercomputability of
analogue machines, for a discussion on this see~\cite{kieu-mindmach}
and references therein.}.
%

In general, we have the following theorem, a proof of which is
available in~\cite{CaludeBookRandom},
\begin{thm}[Finitely Refutable Problems]
Let $s\in\mathbb N$ and $P$ be a $k$-ary predicate on $\mathbb N$,
then every formula
\begin{eqnarray}
f &=& Q_1n_1\, Q_2n_2\cdots Q_sn_s\, P(n_1, n_2, \cdots, n_s),
\end{eqnarray}
where $Q_1, Q_2, \cdots Q_s$ are alternating quantifier symbols, is
finitely solvable.

That is, for each formula $f$ of the form above there exists a
finite set $\T_f \subset \mathbb N^s$ such that: {\bf $f$ is true in
$\mathbb N^s$ iff it is true in $\T_f$}.
\end{thm}
It is the allowing of non-constructive arguments in proofs that
leads to the above phenomenon of such sets $\T_f$ being finite but
can be undecidable. We will provide below some other simpler
arguments for the finitely refutable character of Hilbert's tenth
problem in particular, and discuss how a single procedure in quantum
adiabatic computation could determine, in a probabilistic manner,
the different finite domain for each and every Diophantine equation.

\section{Quantum Adiabatic Algorithm for Hilbert's tenth problem}
We first introduce the Fock occupation-number states $|n\rangle$, $n
= 0, 1, 2, \ldots$, and the creation and annihilation operators
$a^\dagger$ and $a$ respectively
\begin{eqnarray}
a |0\rangle &=& 0,\nonumber\\
a |n\rangle &=& \sqrt{n}\,|n-1\rangle,\nonumber\\
a^\dagger |n\rangle &=& \sqrt{n+1}\,|n+1\rangle,\nonumber\\
(a^\dagger a) |n\rangle &=& n\,|n\rangle,
\label{SHOoperators}
\end{eqnarray}
such that the operators satisfy the commutation relation
\begin{eqnarray}
[a, a^\dagger] &=& 1.
\end{eqnarray}
We also need the definition of coherent state, with a complex number
$\alpha$,
\begin{eqnarray}
|\alpha\rangle &=& e^{-\frac{|\alpha|^2}{2}}\sum_{n=0}^{\infty}
\frac{\alpha^n}{\sqrt{n!}}\,|n\rangle. \label{coherent}
\end{eqnarray}

\subsection{An observation}
Given a Diophantine equation with $K$ unknowns,
\begin{eqnarray}
D(x_1,\ldots,x_K)=0, \label{equation}
\end{eqnarray}
the whole point of the algorithm to be presented below is to obtain
the ground state $|g\rangle$ of a specific quantum mechanical
Hamiltonian,
\begin{eqnarray}
H_P &=& \left(D(a^\dagger_1 a_1, \ldots, a^\dagger_K a_K)\right)^2,
\label{final}
\end{eqnarray}
which is constructed in such a way to reflect the input Diophantine
polynomial. We assume here that the ground state is non-degenerate,
and will discuss the case of degeneracy later. This ground state is
a Fock state, with a particular $K$-tuple $(n^0_1, \cdots, n^0_K)$,
\begin{eqnarray}
|g\rangle &=& |\{n\}^0\rangle = \bigotimes_{i=1}^K |n^0_i\rangle,
\end{eqnarray}
and is associated with the lowest energy of $H_P$, which is also
{\em the global minimum} of the square of the Diophantine polynomial
in the domain of non-negative integers,
\begin{eqnarray}
E_g &=& (D(n^0_1, \cdots,n^0_K))^2, \nonumber\\
&=& \min_{\{n\}}\,(D(n_1, \cdots,n_K))^2 \ge 0.
\label{globalminimum}
\end{eqnarray}
Clearly, the knowledge of $|g\rangle$ enables us to draw a
conclusion about the existence, or not, of solution
for~(\ref{equation}) either directly through its ground-state energy
$E_g$ or indirectly through the occupation numbers $(n^0_1, \cdots,
n^0_K)$, which can then be substituted into the polynomial. Only
when $E_g$ or the substitution vanishes that the equation has any
solution.

The lesson then is, instead of looking for the zeros of any given
Diophantine equation, which may not exist, we will search for the
absolute minimum of the square of the Diophantine polynomial, {\em
which is finite and always exists at a finite point in its domain}.

\subsection{Quantum Adiabatic Computation}
In general, it is much more difficult to construct a specific state
for a quantum mechanical system than to control the physical process
(that is, to create a corresponding Hamiltonian) to which the system
is subject. One systematic method to obtain the ground state of a
Hamiltonian is to exploit the quantum adiabatic theorem to reach the
desired state through some adiabatic evolution which starts from a
readily constructible ground state of some other Hamiltonian. This
is the idea of quantum adiabatic computation (QAC)~\cite{qac}, which
is an alternative to the more standard model of quantum computation
based on qubits and quantum gates in quantum networks~\cite{qcbook}.

In QAC, we encode the solution of our problem to the ground state of
some specific Hamiltonian. As it is easier to implement controlled
dynamical processes than to obtain the ground state, we start the
computation with the system prepared in a different but readily
obtainable ground state of some other Hamiltonian. This initial
Hamiltonian is then slowly extrapolated into the Hamiltonian whose
ground state is the desired one. The adiabatic theorem of quantum
mechanics (QAT)~\cite{messiah} stipulates that if the extrapolation
rate is sufficiently slow compared to some intrinsic scale, the
initial state will evolve into the desired ground state with a high
probability -- the slower the rate of change, the higher the
probability in the adiabatic regime. Measurements then take place
finally on the system in order to identify the ground state, from
which the solution to our problem emerges.

In comparison to the standard quantum computation networks of
unitary gates acting sequentially on a system of qubits, QAC might
be implemented with a continuous computation time as opposed to the
discrete time steps of the former. For finite-dimension Hilbert
spaces QAC may be approximated by the standard quantum computation
networks with an appropriate number of qubits~\cite{Aharonov04}.
However, with QAC at its most generality we are not necessarily
restricted to using qubits or even finite-dimension Hilbert spaces.
As a result of the possibility of applying QAC in dimensionally
infinite or unbounded spaces, QAC could be more than just an
alternative quantum computation model, but could extend the notion
of computability {\em beyond} classical computability, as will be
shown below.

Now, to carry out a QAC for a given Diophantine
equation~(\ref{equation}), we prepare our quantum mechanical system
in the readily constructible initial ground state
\begin{eqnarray}
|g_I\rangle &=& |\{\alpha\}_I\rangle \equiv \bigotimes_{i=1}^K
|\alpha_i\rangle, \label{initial state}
\end{eqnarray}
of a {\em universal} (that is, independent of the given Diophantine
equation) initial Hamiltonian $H_I$, with some complex numbers
$\alpha$'s,
\begin{eqnarray}
H_I &=& \sum_{i=1}^K(a^\dagger_i - \alpha_i^*)(a_i - \alpha_i).
\label{initial}
\end{eqnarray}
This is just the Hamiltonian for shifted simple harmonic oscillators
whose ground state is the well-known coherent state in quantum
optics. We then subject the system to the time-dependent Hamiltonian
$\h(t)$, which linearly extrapolates the initial Hamiltonian $H_I$
to the final Hamiltonian $H_P$ in a time interval $T$,
\begin{eqnarray}
\h(t) &=& \left(1 -\frac{t}{T}\right)H_I + \frac{t}{T}\,H_P.
\label{thehamiltonian}
\end{eqnarray}

\subsection{Statement of the algorithm}
The following quantum adiabatic algorithm will decide in finite time
whether equation~(\ref{equation}) has any non-negative integer
solution or not -- {\em to within certain probabilistic error, which
can be specified arbitrarily and in advance, characterised by the
pair of real numbers $(\epsilon, \delta)$} (see Subsection~\ref{prob
nature} below):
\begin{itemize}
\item Construct/simulate a physical process in which a system initially
starts in a state that is a direct product of $K$ coherent states
(with arbitrarily chosen $\alpha$'s non-zero):
\begin{eqnarray}
|\psi(0)\rangle &=& |\{\alpha\}_I\rangle, 
\end{eqnarray}
and in which the system is subject to a time-dependent Hamiltonian
$\h(t)$ of ~(\ref{thehamiltonian}) over the time interval $[0,T]$,
for {\em some} time $T$.

\item Measure (or calculate through the Schr\"odinger equation with the
time-dependent Hamiltonian above, see eq. (\ref{Schrodinger})) the
maximum probability to find the system in a particular
occupation-number state at the chosen time $T$,
\begin{eqnarray}
P(T) &=& \max_{|\{n\}\rangle}\,|\langle\psi(T)|\{n\}\rangle|^2,\nonumber\\
&=&\left|\langle\psi(T)|\{n\}^0\rangle\right|^2, \label{probability}
\end{eqnarray}
where $|\{n\}\rangle = \bigotimes_{i=1}^K|n_i\rangle$, and
$|\{n\}^0\rangle$ is the maximum-probability number state with a
particular $K$-tuple $(n^0_1, \cdots, n^0_K)$.
\item If $P(T)\le 1/2$, increase $T$ and repeat all the steps above.
\item If $P(T)>1/2$ (to within some arbitrarily predetermined uncertainty
$(\epsilon,\delta)$, see later) then the algorithm can be terminated
as $|\{n\}^0\rangle$ is the ground state of $H_P$ (assuming no
degeneracy, we discuss the case of degeneracy below), and a
conclusion can now be deduced from the fact that: {\em The
equation}~(\ref{equation}) {\em has a non-negative integer solution
iff $H_P|\{n\}^0\rangle = 0$}.
\end{itemize}
We refer to~\cite{kieu-contphys, kieu-intjtheo, kieu-royal} for the
motivations and discussions leading to the algorithm above. Here, we
would like to make some immediate remarks.

Note that it is crucial that $H_I$ does not commute with $H_P$,
\begin{eqnarray}
[H_I,H_P] &\not =& 0, \label{commutation}
\end{eqnarray}
and of course both Hamiltonians are of infinite dimensions. (This
condition can be relaxed to that of unbounded dimensions, see
later.)

To remove any possible degeneracy of the ground state for any $H_P$,
we can always introduce to $H_P$ a symmetry-breaking term of the
form $(\gamma a^\dagger_j+\gamma^* a_j)$ for some
$j$~\cite{kieu-royal}. This term destroys the symmetry generated
from the commutation between $H_P$ and the occupation-number
operators, $a^\dagger_j a_j$. However, we can recover the symmetry
in the limit in the limit ${|\gamma|}\to 0$, and modify the
algorithm above slightly to reach an answer for the Diophantine
equation in question. Note also that our arguments for the
non-degeneracy of the ground state of $\h(t)$ in $0\le t <
T$~\cite{kieuFull} can be easily modified to accommodate this
symmetry-breaking term. From here on we will discuss only the case
of no degeneracy in the ground state of $H_P$.

In the rest of this Section on the algorithm, we will discuss its
other crucial and fundamental aspects in the following Subsections.

\subsection{Quantum Adiabatic Theorem}
QAC relies crucially on the theorem of quantum adiabatic processes,
which has been experimentally confirmed and mathematically proven
not only for dimensionally finite but also for dimensionally
infinite or unbounded spaces~\cite{messiah}.
\begin{thm}[Quantum Adiabatic Theorem]
Let $H(s)$ be a Hamiltonian in the interval $0\le s \le 1$:
\begin{enumerate}
\item Having discrete eigenvalues $\epsilon_i(s)$ which, together
with the projectors $\mathbb P_i(s) = |\epsilon_i(s)\rangle\langle
\epsilon_i(s)|$ onto their respective subspaces, are continuous
functions of $s$;
\item Having non-degenerate eigenvalues throughout the interval:
$\epsilon_i(s) \not = \epsilon_j(s)$, for $i\not = j$;
\item Having the derivatives $\dd\mathbb P_i/\dd s$ and $\dd^2\mathbb
P_i/\dd s^2$ being well-defined and piece-wise continuous in the
whole interval;
\end{enumerate}
then
\begin{eqnarray}
\lim_{T\to\infty} U_T(s)\,\mathbb P_i(0) &=& \mathbb
P_i(s)\lim_{T\to\infty}U_T(s),
\end{eqnarray}
where $U_T(s)$ is the time evolution operator generated by the
time-dependent $H(s)$,
\begin{eqnarray}
i\hbar\frac{\dd}{\dd s}\,U_T(s) &=& T\,H(s)\,U_T(s),\\
U_T(0) &=& \bf 1.\nonumber
\end{eqnarray}
\end{thm}
In other words, if the system is initially in the $i$-th eigenstate
then it remains in the same instantaneous eigenstate, up to a phase
$\phi$ and provided the adiabaticity conditions are satisfied,
\begin{eqnarray}
\lim_{T\to\infty} U_T(s)\,|\epsilon_i(0)\rangle &=& {\rm
e}^{i\phi(s)}|\epsilon_i(s)\rangle.
\end{eqnarray}
Here, the inverse of $T$ is a measure of the rate of change of the
time-dependent Hamiltonian.

To apply this theorem for QAC we need to satisfy its assumptions
first of all. In particular, we have shown elsewhere~\cite{kieuFull}
the crucial condition of no degeneracy of (that is, no level
crossing for) the instantaneous ground state of interest of the
time-dependent Hamiltonian~(\ref{thehamiltonian}) in the open
interval $0\le t<T$, with~(\ref{initial}) and~(\ref{final}) being
respectively the initial and final Hamiltonians. The proof is quite
involved and it suffices to point out here that we first establish
the non-degeneracy for real and positive $\alpha_i$'s
of~(\ref{initial}) by employing some theorem for dimensionally
infinite but {\em bounded} operators~\footnote{Theorem XIII.44
in~\cite{ReedSimon}, pp. 204--205. A cut-down version of which can
be rephrased for our purpose: ${\h}$ has a non-degenerate ground
state iff ${\rm e}^{-a{\h}}$ is {\em positive improving} for all
$a>0$.} We can in fact prove a stronger result, by invoking another
theorem of Reed and Simon~\footnote{Theorem XIII.43
in~\cite{ReedSimon}, pp. 202--204.}, that {\em all} the eigenvectors
of ${\h}(t)$, not just the ground state, are non-degenerate for real
and positive $\alpha_i$'s and for $0\le t< T$. These results are
then generalisable to complex-valued $\alpha_i$'s thanks to a gauge
symmetry of the full
Hamiltonian~(\ref{thehamiltonian})~\footnote{All these also agree
with our alternative arguments in~\cite{kieu-royal} which lead to
the non-degeneracy of the ground state of ${\h}(t)$ for $0\le
t<T$.}. The possible degeneracy of $\h(T) = H_P$ can then be handled
as discussed in a previous Subsection.

It thus follows from QAT that if the instantaneous ground state
never crosses with any other state in the spectral flow then it
takes a non-vanishing rate of change $1/T$, and thus only a {\em
finite} time $T$, to obtain the final ground state with a
probability which could be made arbitrarily close to one. {\em This
implies that our quantum algorithm can always be terminated in a
finite time.}

Nevertheless, it is important to note that QAT is {\em not
constructive}, as with most theorems involving limiting processes, .
It only tells us that for ``sufficiently large" $T$ the system is
``mostly" in the instantaneous eigenstate. But the theorem tells us
nothing quantitatively about the degrees of being ``sufficiently
large" or ``mostly". This is the general result of the fact that a
limit is often stated in the language that given any $\epsilon > 0$
then there exists a $\delta(\epsilon) > 0$, as an {\em unspecified}
function of $\epsilon$ in general, such that certain condition
involving $\delta$ has to be satisfied first as a mathematically
sufficient condition for a desirable statement involving $\epsilon$.

In other words, QAT can only guarantee that the ground state is
achievable in a finite time interval but cannot specify what that
interval should be. That is, it cannot by itself give us any
indication when the ground state has been obtained so that the
algorithm can then be terminated at that point. For that, we need
another criterion and the next Subsection is devoted to this
criterion.

\subsection{Identification of the ground state and termination of the algorithm}
The crucial step of any quantum adiabatic algorithm is the
identification of the ground state of the final Hamiltonian, $H_P$.
In our case we do not in advance know in general how long is
sufficiently long (the Theorem offers no direct help here); all we
can confidently know is that for each Diophantine equation and each
set of $\alpha_i$'s there is a {\em finite} evolution time after
which the adiabaticity condition is satisfied. We thus have to find
another criterion to identify the ground state.

The identification criterion we have found can be stated as:
\begin{prop}[Identification of the Ground State]
The ground state of $H_P$ is the Fock state $|\{n\}^0\rangle$
measuredly obtained with a probability of more than ${1}/{2}$ after
the evolution for some time $T$ of the initial ground state
$|g_I\rangle$ according to the Hamiltonian~(\ref{thehamiltonian}):
\begin{center}
$|\{n\}^0\rangle$ is the ground state of $H_P$ {\bf iff} 
$\left|\langle\psi(T)|\{n\}^0\rangle\right|^2 > {1}/{2}$, for some
$T$,
\end{center}
provided the initial ground state $|g_I\rangle$ of $H_I$ does not
have any dominant component in the occupation-number eigenstates
$|\{n\}\rangle$ of $H_P$,
\begin{eqnarray}
|\langle g_I|\{n\}\rangle|^2 \le {1}/{2},\forall\, \{n\}.
\label{nodominant}
\end{eqnarray}
\label{id}
\end{prop}
Note that the condition~(\ref{nodominant}) is crucial
here~\footnote{If it does not hold then we can use the sudden
approximation to show that there exists some range of $T$ such that
$\left|\langle\psi(T)|\{m\}\rangle\right|^2 > {1}/{2}$ but
$|\{m\}\rangle$ is not the ground state of $H_P$.}. And our choice
of the coherent state~(\ref{initial state}) as the initial ground
state entails that the condition~(\ref{nodominant}) is always
satisfied, since for any $\alpha\not=0$ and $\forall n$:
\begin{eqnarray}
|\langle \alpha|n\rangle|^2 &=& e^{-|\alpha|^2}{|\alpha|^{2n}}/{n!}
< {1}/{2}.
\end{eqnarray}
And for $K$-tuples, the bound is even better satisfied:
\begin{eqnarray}
|\langle \{\alpha\}|\{n\}\rangle|^2 &<& \left({1}/{2}\right)^K.
\label{k-tuple}
\end{eqnarray}

We can recognise QAT in the {\em `only if'} part of the statement
above in Proposition~\ref{id}; and we only need to prove the {\em
`if'} part. We have first proved this criterion for two-state
systems~\cite{kieuFull} and then argued that it is also applicable
for systems of finitely many and of infinite number of states
because only two states, the instantaneous ground state and first
excited state, among all the available states become dominantly
relevant at any instant of time -- provided we start out with the
ground state of the initial Hamiltonian $H_I$. We have now been able
to prove this result for an infinite number of levels. The proofs
are non-constructive, and their conclusions have also been
numerically confirmed in several simulations of our
algorithm~\cite{kieu-spie} and also of a modified
version~\cite{columbia2, columbia}.

Without going into the details, to derive the result above we first
show that the probability of finding $|\psi(T)\rangle$, starting
from the ground state of $H_I$ at $t=0$, to be in {\em any excited
state} (that is, not the ground state) of $H_P$ is always less than
$1/2$, for any duration $T$. Then combining with QAT, which dictates
that there must exist a measuredly obtained state contributing
to~(\ref{probability}) such that $\lim_{T\to\infty} P(T) = 1$,
hence: (i) there must exist a critical time $T_c$ beyond which we
have $P(T)> 1/2$; and (ii) such a state then has to be the ground
state of $H_P$.

Note that both the proof and result of the
Proposition~\ref{nodominant} are non-constructive. The result is
non-constructive because it does not tell us what the time $T_c$ is.
It cannot, simply because such a time is specific to each different
Diophantine equation, and there is no finite upper bound for {\em
all} Diophantine equations. But that is fine because QAT asserts
that such a time $T$ {\em is finite for each individual equation},
even though QAT cannot say anything in general about $P(T)$ as a
function of $T$.

Now, combining QAT with the above Proposition, we only need to
increase the evolution time $T$ until one of the occupation-number
states is measuredly obtained at time $T$ with a probability more
than ${1}/{2}$. This will be our much desired ground state -- to
within certain probabilistic error characterised by the pair of real
numbers $(\epsilon, \delta)$, see below. Then and only then, we
terminate the algorithm~\footnote{Note that since we are looking for
a state which has a measurement probability more than even for some
evolution time $T$, we can employ the boosting technique discussed
in~\cite{bernsteinetal, kieu-intjtheo} to find the ground state by
the majority rule in performing the computation over $l$
concatenated copies of the underlying Hilbert space, instead just
over the original Hilbert space. The probability of finding our
ground state by the majority rule (that is, if more than $l/2$
states obtained in the direct product of $l$ states are the same
state then this common state in the direct product is the ground
state we are after) could be boosted to a value arbitrarily close to
unity $(1-\epsilon')$, without the need of further increasing $T$,
provided that $l > - C(T) \log\epsilon'$ with $C$ is some function
of $T$.}.

\subsection{\label{prob nature}Probabilistically correct nature
of the algorithm}
It is important to recognise that {\em the findings of our algorithm
are subject to probabilistic errors -- that is, they have some
pre-specified but non-zero probabilities of being wrong!}

As the halting criterion, the probability $P(T)$
of~(\ref{probability}) needs to be more than one-half. If the
algorithm is implemented by certain physical process then we can
approximate this probability by the relative frequency $\nu_N$ that
a particular state is obtained in $N$ repetitions. The Weak Law of
Large Numbers, however, can only assert that this frequency $\nu_N$
is within a distance $\epsilon$ of $P(T)$ {\em with a probability}
of at least $(1-\delta)$.
\begin{thm}[The Weak Law of Large Numbers]
Let
\begin{eqnarray}
\nu_N &=& \frac{1}{N}\sum_{i=1}^N \xi_i,
\end{eqnarray}
where $\xi_i$, $i=1,\cdots,N$, are independent identically
distributed random variables with mean $p$ and variance $\sigma$.
Then $\forall \,\epsilon > 0$ and $\forall\, \delta >0$,
\begin{eqnarray}
N > \frac{\sigma^2}{\epsilon^2\delta} &\Longrightarrow&
P\left\{\left|\,\nu_N - p\, \right| > \epsilon \right\} \le \delta.
\label{weak}
\end{eqnarray}
\label{WeakLaw}
\end{thm}
See, for example,~\cite{probtheory} for a proof of the theorem.
Clearly, if we let $\xi_i=1$ if $|\{n\}^0\rangle$ is obtained in the
$i$-th measurement, and $\xi_i=0$ otherwise, then $\nu_N$ is the
relative frequency in all $N$ measurements. Also then, $p$
in~(\ref{weak}) is our $P(T)$ of~(\ref{probability}), and $\sigma^2$
(which equals to $\langle \xi^2\rangle - \langle \xi\rangle^2$) is
$p\,(1-p\,)$ and cannot be more than a quarter.

It is the probability of error $\delta$ in~(\ref{weak}) that gives
the quantum algorithm its probabilistically correct nature. Both
$\epsilon$ and $\delta$ are dependent on the number of measurement
repetitions from which the frequency $\nu_N$ is obtained. In
general, we can always reduce $\epsilon$ and $\delta$ to arbitrarily
close to zero by increasing the number of repetitions $N$
appropriately,
\begin{eqnarray}
N > 1/(4\epsilon^2\delta). \label{L}
\end{eqnarray}
In this instance, we have been able to obtain $N$ {\em
constructively} as a function of $\epsilon$ and $\delta$ from the
Weak Law of Large Numbers~\ref{WeakLaw}, which is {\em
non-constructive} in general since a bound on $\sigma$ there is not
always known.

{\em For the quantum adiabatic algorithm of this paper, the pair
$(\epsilon, \delta)$, which can be specified arbitrarily and in
advance, characterises the probabilistic uncertainty of the findings
of the algorithm.}

We give this kind of algorithms the name probabilistically correct
algorithms to emphasise the fact that not only their computation
steps follow certain probabilistic evolutions but the end results
from such algorithms are subject to some {\em probabilities of being
incorrect}. Such error probabilities are necessary when there is, in
principle, no other way to verify all the outcomes of the
algorithms. Nevertheless, the value of this class of algorithms lie
in the fact that the error probabilities are not only known but
could also be predetermined with arbitrarily small values.
\begin{defn}[Probabilistically Correct Algorithms] \em A probabilistic
correct computation is a probabilistic computation whose outcome has
a correctness which can be pre-assigned with an arbitrary
probability. \label{PCC}
\end{defn}
Another example of this kind of probabilistically correct
computations is the computation supplied with biased coins as
oracles as discussed in~\cite{OrdKieu-qubit}. Probabilistically
correct algorithms should be distinguished from other (and more
common) probabilistic algorithms whose outcomes can be verified in
principle (by some other means) and thus be confirmed without any
uncertainty.

As an anticipation of Section~\ref{numerical} below, we mention now
that the probabilistically correct nature of the algorithm is
manifest differently in the numerical simulations of the quantum
adiabatic algorithm there through the uncertainty associated with
the extrapolation to zero step size in solving the Schr\"odinger
equation.

\subsection{\label{outside}Outside the jurisdiction of Cantor's arguments}
In claiming that our quantum algorithm can somehow compute the
noncomputable, we also need to consider it in the context of the
no-go arguments of Cantor in Subsection~\ref{no-go}. Those
arguments, indeed, cannot be applicable here for several reasons.
First of all, the proof for the working of our algorithm is not
constructive, implying its non-recursiveness. The mathematical
proofs for QAT itself and for the criterion for ground-state
identification~\cite{kieuFull} are highly non-constructive; also
have to be employed at several places are the methods of proof by
contradiction and also of (non-constructive) analysis for continuous
functions.

Secondly, and more explicitly, our algorithm is outside the
jurisdiction of those no-go arguments because of its probabilistic
nature. We have argued elsewhere~\cite{OrdKieu-qubit} (see
also~\cite{shannon, Santos:1971}) against the common
misunderstanding that probabilistic computation is equivalent in
terms of computability to Turing recursive computation. They are
not! We have pointed out that, for example, if a non-recursively
biased coin is used as an oracle for a computation, the computation
carries more computability than Turing computation in general.

Here, we now explicitly show how the probabilistic nature of our
algorithm can avoid the Cantor's diagonal no-go
arguments~\footnote{The seed of the following arguments was
originated from a group discussion with Enrico Deotto, Ed Farhi,
Jeff Goldstone and Sam Gutmann in 2002. It goes without saying that
if there are mistakes in the interpretation herein, they are solely
mine.}.

Because our algorithm is probabilistically correct, we can only
obtain an answer with certain probability to be the correct answer.
This probability can be made arbitrarily close, but never equal, to
one as shown in~(\ref{L}). Thus, instead of the halting
function~(\ref{Cantor}) our algorithm can only yield a probabilistic
halting function $ph$, which must have {\em three} arguments instead
of two,
\begin{eqnarray}
\\
ph(p,i,\delta)&=&\left\{
\begin{tabular}{l}
0 \mbox{ if $p$ halts on input $i$, with maximun error-probability $\delta$};\\
1 \mbox{ if does not halt, with maximun error-probability $\delta$}.
\end{tabular}
\right. \label{pCantor} \nonumber
\end{eqnarray}
Following the flow of Cantor's arguments, one can then construct a
program $pTuring(n,\delta)$ having arguments $n$ and $\delta$ in
such a way that it calls the function $ph(n,n,\delta)$ as a
subroutine and then halts if and only if $ph(n,n,\delta) = 1$. In
some made-up language:
\begin{quotation}
\begin{tabular}{ll}
\tt
 &\tt Program $pTuring$\\
 &\tt input ($n$,$\delta$)\\
\tt 10& \tt call $ph(n,n,\delta)$\\
 &\tt if $ph(n,n,\delta) = 0$ goto 10\\
 &\tt stop\\
 &\tt end
\end{tabular}
\end{quotation}
Similarly, let $t_p$ be the G\"odel encoded integer for the program
$pTuring$~\footnote{Barring the case $pTuring$ does not have an
integer encoding, which is quite possible for a quantum
algorithm~\cite{kieu-intjtheo}, in that case the Cantor's diagonal
arguments cannot be applied anyway.}. We next apply the
probabilistic halting function $ph$, with some $\delta'$, to $t_p$
and $(n,\delta)$, which is the total input for $pTuring$, to obtain:
\begin{quotation}
$ph(t_p,\tilde{n},\delta') = 0$ {\bf iff } $ph(n,n,\delta) = 1$,
\end{quotation}
where $\tilde{n}$ encodes $(n,\delta)$ uniquely~\footnote{For
example, for $\delta\not=0$ and $\delta = 2^{-J}$, one could use the
encoding $\tilde n = n + J$, or $\tilde n = p_1\,^np_2\,^J$ where
$p_1$ and $p_2$ are two different prime numbers.}. No matter what we
choose for $n$ and $\delta'$ and $\delta$, we {\em cannot Cantor
diagonalise} the above because $\tilde n$ can never be equated to
$n$:
\begin{itemize}
\item because neither $\delta$ nor $\delta'$ can be
zero, unlike the previous situation of definitely correct
computational results in Subsection~\ref{T1} (where both $\delta$
and $\delta'$ vanish);
\item because neither $\delta$ nor $\delta'$ can be one, as we would
then have definitely incorrect computational results, from which we
could have deduced the definitely correct answer. (For decision
problems, where the solutions can only be ``yes" or ``no", knowing
that the computational results are definitely incorrect would enable
us to reach definitely correct answers -- simply by taking the
negation of the computational outcomes.)
\end{itemize}
Thus we cannot use any mathematical contradiction or inconsistency
here to dismiss the existence of the function $ph$.

The above arguments clearly have nothing to do with being quantum
mechanical, but everything to do with being probabilistic. Quantum
mechanics, however, has given us the inspiration and the means to
realise a probabilistically correct algorithm capable of deciding
Hilbert's tenth problem with a single, universal procedure for any
input of Diophantine equations.
\begin{prop}[The Power of Probabilistically Correct Computation] The
set of Turing computable problems is a strict subset of the set of
probabilistically correctly computable problems.
\end{prop}
The proof follows from the above and from the fact that Turing
computation is equivalent to the  probabilistic
computation~\cite{shannon} whose computational branching probability
is a computable number and whose outcomes can be verified with zero
uncertainty.  This latter type of computation in turn belongs to the
more general probabilistically correct computation as in
Definition~\ref{PCC}.

\subsection{Finite but unbounded number of dimensions}
Our quantum algorithm seems to be an infinite search through the
positive integers in a finite time! This may sound strange unless
one recognises that:
\begin{enumerate}
\item The {\em absolute} minimum of the square of any Diophantine
polynomial {\em always} takes place at some tuple of finitely valued
nonnegtive integers. This is also a manifestation of the finitely
refutable property discussed in Subsection~\ref{finitely} for
Hilbert's tenth problem, because if the Diophantine equation has no
solution up to that tuple of finite integers then it cannot have any
solution in the whole domain of non-negative integers.
\item And even though in a finite time the wavefunction starting
from an initial ground state can only cover essentially a finite
domain of the Fock space, the finite domain so covered is indeed the
relevant domain which encloses the point at which the absolute
minimum is obtained. That is, the initial ground state provides one
end of the thread along which we can trace to the final ground state
at the other end.
\end{enumerate}
In this way, we see that, contrary to naive expectation, we may not
need an infinite space to carry out the algorithm for any given
Diophantine equation. A sufficiently large space will do, since once
the final ground state has already been included in this space the
addition of further Fock states which are highly excited states of
$H_P$ will only change the quantum dynamics negligibly. This is once
again a direct result of the quantum adiabatic theorem which tells
us that in the adiabatic regime only the ground state and first few
excited states of $\h(t)$ are dominantly relevant to the dynamics.

Obviously, the key point is how large is {\em sufficiently large}
for a given Diophantine equation. And of course this cannot be
determined recursively in general by a single procedure, otherwise
Hilbert's tenth problem would have been Turing computable.
Nevertheless, we argue that the existence of universal means which
can {\em probabilistically} determine when the underlying Fock space
is sufficiently large cannot be ruled out. Such means would also
acceptable since our quantum algorithm is only probabilistically
correct anyway -- provided we could find suitable probability
measures to quantify the degrees of sufficiency of such truncations
to finite Fock spaces. In a sense, the number of dimensions of the
underlying Fock space could be the same as the length of the tape
(finite but unbounded) of a Turing machine; we do not need an
infinitely long tape to start a Turing computation but can lengthen
the tape as needed as the computation proceeds.

In the Section~\ref{numerical} below we present some numerical
simulation results for the quantum processes~\cite{kieu-spie} for
some simple Diophantine equations, and demonstrate (with an
incomplete procedure) how to find the number of dimensions for the
problems under consideration.

\section{\label{numerical}Some numerical simulations~\cite{kieu-spie}}
\subsection{The approach}
The time evolution of the algorithm is governed by the Schr\"odinger
equation
\begin{eqnarray}
i\frac{\partial}{\partial t}\, |\psi(t)\rangle &=&
\h(t)\,|\psi(t)\rangle, \;{\mbox{\rm for }}0<t<T,\; {\mbox{\rm with
}}|\psi(0) = |\alpha_I\rangle, \label{Schrodinger}
\end{eqnarray}
from which $P(T)$ of~(\ref{probability}) could be computed in order
to identify the deciding final ground state of $H_P$.

Now, based on the fact that Hilbert's tenth problem is not
recursively computable, we have to conclude that the class of linear
differential equations above are not Turing computable. This is new
to the class of ordinary three-dimensional wave equations that have
been found by Pour-El and Richards to be
non-computable~\cite{PourEl}. We have briefly
commented~\cite{kieu-royal} on this noncomputable behaviour of these
Schr\"odinger equations~(\ref{Schrodinger}) in the context of a
non-computability theorem by Pour-El and Richards, which deals with
bounded linear operators on some computability structures. We hope
to develop the connection further elsewhere, but it suffices to
identify here the noncomputability of the Schr\"odinger equations is
due to the fact that we cannot numerically implement a dimensionally
infinite Hilbert space.

Notwithstanding this, the Schr\"odinger equations and hence the
algorithm should require only sufficiently large but finite Hilbert
space with a truncated basis consisting of occupation-number states,
$\{|n\rangle: n=0,1,\ldots, m\}$, for some $m$ specific to each
problem, provided that the final ground state of $H_P$ has already
been included therein. The higher occupation-number states beyond
$m$ are expected to contribute only negligibly to the dynamics of
the energetically low-lying states relevant to our problem. This
follows from the facts that the normalised wavefunction at any given
time has a support which spreads significantly only over a finite
range of occupation-number states, and that the explicit coupling,
and thus the influence, between one particular instantaneous
eigenstate and others diminishes significantly outside some finite
range  (as can be seen through a set of differential equations
in~\cite{kieu-royal} connecting instantaneous eigenstates of
${\h}(t)$). Also thanks to QAT, in the adiabatic regime only the
energetically low-lying states are relevant and further inclusion of
higher excited states will not change the ordering of the low-lying
states and affect the dynamics only negligibly. We exploit these
facts fully in the numerical simulations of the quantum
algorithm~\cite{kieu-spie}:
\begin{itemize}
\item We solve the Schr\"odinger equation~(\ref{Schrodinger})
numerically in some finitely truncated Fock space large enough to
approximate the initial coherence state to an arbitrarily given
accuracy. That is, a truncation $m$ is chosen such that our initial
state
\begin{eqnarray}
|\{\alpha\}_I;m\rangle &=& \bigotimes_{j=1}^K |\alpha_j;m\rangle,
\end{eqnarray}
where
\begin{eqnarray} |\alpha_j;m\rangle &=&
e^{-\frac{1}{2}|\alpha_j|^2} \sum_{n=0}^m
\frac{\alpha_j^n}{\sqrt{n!}}\,|n\rangle,\nonumber
\end{eqnarray}
now has a norm less than unity by some predetermined
$\tilde{\epsilon}$,
\[|1- \sqrt{\langle
\{\alpha\}_I;m|\{\alpha\}_{I};m\rangle} | \le\tilde{\epsilon}.\]

\item At each time step $\delta t$, and up to $O(\delta t^2)$,
\begin{eqnarray}
\psi(t+\delta t) &=& \exp\{-i{{\h}}(t)\,\delta t\} \,\psi(t),\nonumber\\
&=& \left(1 -i{{\h}}(t)\,\delta t -\frac{1}{2} {{\h}}^2(t)\,\delta
t^2\right)\,\psi(t).
\end{eqnarray}
Thus this order only involves up to $\h^2$, which in turn involves
only up to the square of the initial Hamiltoninan
$H_I$~(\ref{initial}). That is, up to this order, there are
maximally only two creation operators, $a^\dagger_k a^\dagger_l$
acting on the rhs of the last equation. Thus, from the
properties~(\ref{SHOoperators}) of creation operators $a^\dagger$,
we see that {\em the underlying truncated Fock space need be
extended only by two occupation-number states (for each of the $k$
index) at each time slice!} This is an important observation and we
can explore this fact to {\it explore the infinite Fock space} {\em
by increasing the size of the truncated Fock space by at least two
(for each $k$ index) at every time slice t}.
\end{itemize}

Here the uncertainty in extrapolating the step size $\delta t$ to
zero (which is equivalent to an extrapolation to an infinite number
of dimensions, as we would require an infinite number of time steps
at zero size to cover any finite time interval) is the counterpart
of the uncertainty in approximating the quantum probabilities by
relative frequencies in some physical implementation as discussed in
Subsection~\ref{prob nature}. Both of these uncertainties should be
quantified by some probability measures. Above in
Subsection~\ref{prob nature}, we have pointed out how the Weak Law
of Large Numbers provides a natural probability measure for the
latter. For the extrapolation of the step size to zero in numerical
simulations, we hope to investigate further and discuss elsewhere if
some suitable probability measures could be introduced. The
numerical examples below do not really require such a probability
measure because they are so simple and the results can be verified
by other known methods. See also other numerical simulations of a
variant of our algorithm~\cite{columbia2, columbia}.

\subsection{Simulation parameters and results}
\subsubsection{Equation $x-20 = 0$} We consider a simple example
which nevertheless has all the interesting ingredients typical for a
general simulation of our quantum algorithm. Here we choose
$\tilde{\epsilon} = 10^{-3}$ and the initial Fock space has only up
to $|n\rangle = |14\rangle$, which does not include the true ground
state of $H_P$. This is typical in our simulations since we in
general would not be able to tell in advance whether our initial
Fock spaces contain the true ground states or not. Generally, they
do not. Even so, our strategy of allowing an expansion in time of
the size of the underlying truncated Fock space has enabled the true
ground state to be found and identified. Here and in all other
simulations below we choose $\alpha_k = (2.0,0.0)$.

\begin{figure}
\begin{center}
\includegraphics{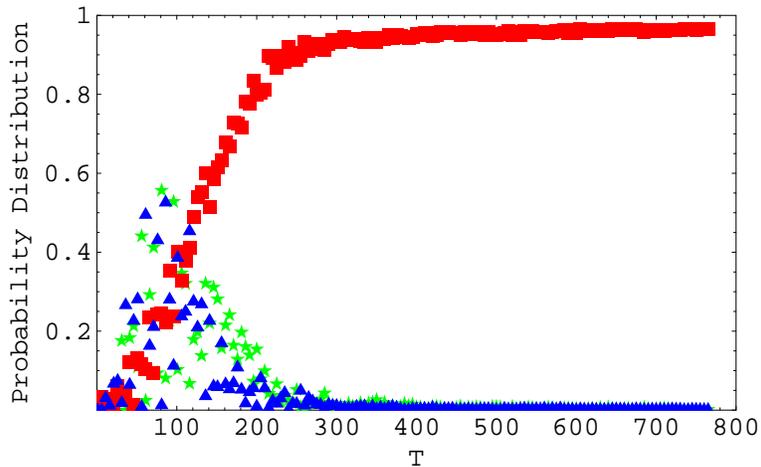}%
\caption{\label{Fig4}Corresponding equation $x-20=0$. See the main
text for details.}
\end{center}
\end{figure}

In Fig.~\ref{Fig4} we plot the magnitude square of the dominant
components (in the basis of Fock states) of the state vector as a
function of the total evolution time $T$ in some arbitrary unit. In
this example, the state $|20\rangle$, {\em which is not included in
the initial truncated Fock space}, is eventually reached by the
expansion of the underlying Fock space by two states at each time
slice. It can be identified as the ground state by the
criterion~(\ref{nodominant}) as shown as (red) boxes in
Fig.~\ref{Fig4}. (Blue) triangles and (green) stars are
corresponding to the first two excited states $|19\rangle$ and
$|21\rangle$, which are degenerate eigenstates of
$H_P$~\footnote{Note that these competing pretenders somehow have
unexpectedly probabilities greater than one-half (around $T\sim
90$), contrary to our analytical result that only the ground state
can have probability rising above one-half! We think that this is
only some artefacts of finite-size time steps $\delta t$, which
should go away once we employ a more sophisticated method for
solving the Schr\"odinger equation.}.

\subsubsection{Equation $xy+x+4y-11=0$}
In this example, contrary to the one above, we fix the truncated
size of our Fock space at all times to be $(m_x,m_y)=(9,9)$ so that
the norm of the state vector is less than unity by an amount
$\tilde{\epsilon} = 10^{-2}$.

\begin{figure}
\begin{center}
\includegraphics{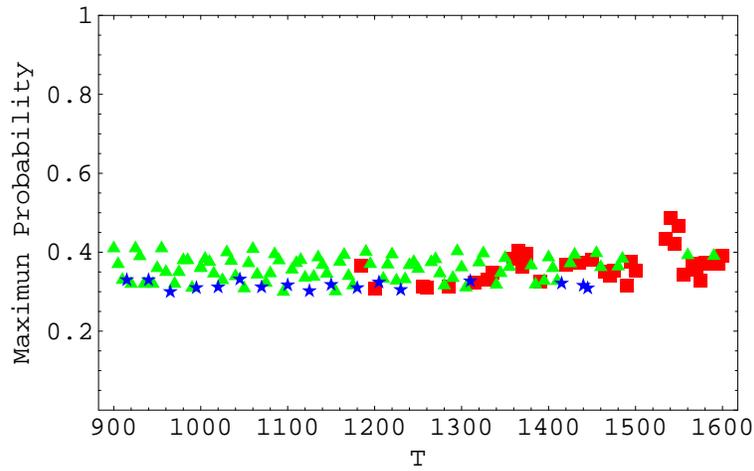}
\caption{\label{Fig1}Corresponding equation $xy+x+4y-11=0$; with
intermediate evolution time $T$. See the main text for details.}
\end{center}
\end{figure}

In Fig.~\ref{Fig1}, below $T\sim 1500$, the maximum probability
components of $|\psi(T)\rangle$ are dominated by some states,
denoted by (blue) star and (green) triangle symbols, none of which
are the actual ground state. In fact, they are, respectively, the
{\em first} degenerate excited states $|n_x, n_y\rangle =
|4,1\rangle$ and $|3,1\rangle$ of $H_P$.

\begin{figure}
\begin{center}
\includegraphics{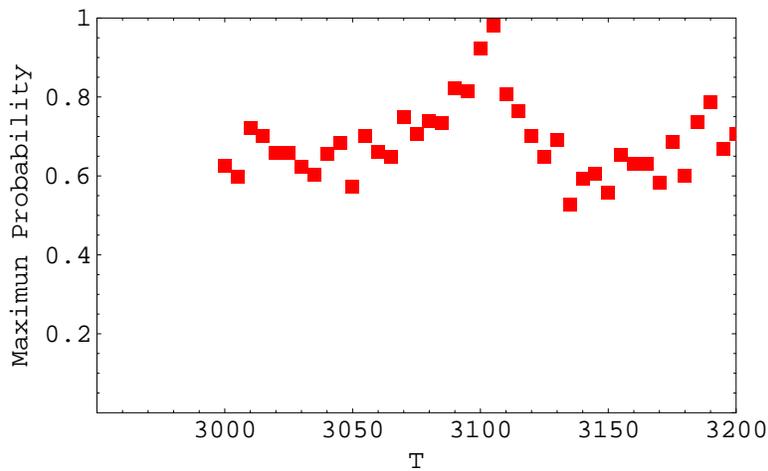}
\caption{\label{Fig2}Corresponding equation $xy+x+4y-11=0$ as in
Fig.~\ref{Fig1} but with $T$ entering the quantum adiabatic regime.}
\end{center}
\end{figure}

With $T$ increased in Fig.~\ref{Fig2}, one of the Fock states,
namely $|1,2\rangle$ denoted by (red) boxes, has the measurement
probability greater than ${1}/{2}$, which is our criterion for being
identified as the ground state. From the ground state so identified
we can infer that our Diophantine equation has a solution, which can
be duly verified.

\subsubsection{Equation $x+20 = 0$} We consider this simple
equation as an example which has no solution in the positive
integers. The simulation parameters are as in the previous, except
that the initial truncated size of our Fock space is up to
$|n\rangle =8$ and is allowed to expand in time.

\begin{figure}
\begin{center}
\includegraphics{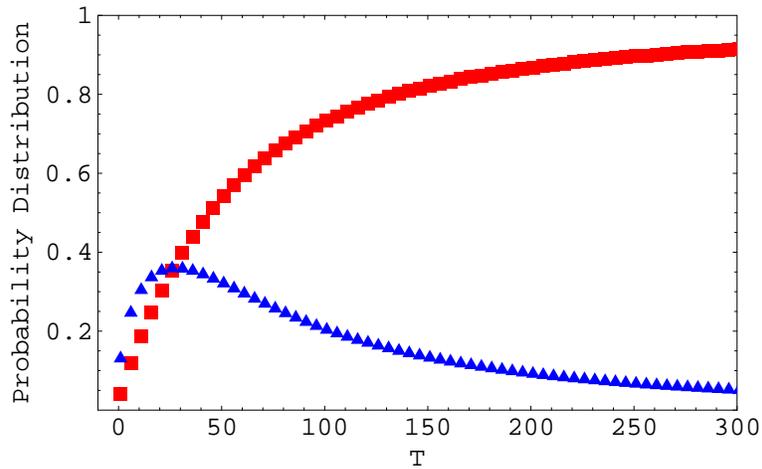}
\caption{\label{Fig3}Corresponding equation $x+20=0$. See the main
text for details.}
\end{center}
\end{figure}

Below $T\sim 50$ in Fig.~\ref{Fig3} none of the two components is
greater than one-half, and in fact the first excited state,
$|1\rangle$ denoted by (blue) triangles, clearly dominates in this
regime. Eventually we enter the quantum adiabatic regime upon when
the (red) boxes rise over the one-half mark; indeed it corresponds
to the Fock state $|0\rangle$, which is the true ground state and
which implies that our original Diophantine equation has no integer
solution at all.

\section{On physical implementations}
\subsection{Misunderstandings and/or prejudices}
Since the first announcement of an earlier version of the quantum
adiabatic algorithm in 2001, there have been numerous discussions in
private but only few postings have yet appeared in the public domain
which are directly related to and/or built on the
algorithm~\cite{Tsirelson, Srikanth, columbia2, columbia,
Srinivasan04, Ziegler04}, see also the Notes added at the end of
this paper.

In this Subsection we will point out some of the misunderstandings,
expressed either privately or publicly, that we know of which are
directed towards the algorithm.
\begin{enumerate}
\item Our proposal is in contrast to the claim in~\cite{bernsteinetal}
that quantum Turing machines compute exactly, albeit perhaps more
efficiently, the same class of functions which can be computed by
classical Turing machines. The quantum Turing machine approach is a
direct generalisation of that of the classical Turing machines but
with qubits and some universal set of one-qubit and two-qubit
unitary gates to build up, step by step, dimensionally larger, but
still dimensionally finite unitary operations. This universal set is
chosen on its ability to evaluate any desirable classical logic
function. Our approach, on the other hand, is already from the start
based on infinite-dimension Hamiltonians acting on some Fock space
and also based on the special properties and unique status of their
ground states. These are the reasons behind the ability to compute,
in a finite number of steps, what the dimensionally finite unitary
operators of the standard quantum Turing computation cannot do in a
finite number of steps.

Indeed, several authors (including Nielsen~\cite{nielsen} and Calude
and Pavlov~\cite{CaludePavlov}) have also found no logical
contradiction in applying the most general quantum mechanical
principles (over and above those employed in the standard quantum
computation) to the computation of the classical noncomputable,
unless certain Hermitean operators cannot somehow be realised as
observables or certain unitary processes cannot somehow be admitted
as quantum dynamics. And up to now we have neither any evidence nor
any principles that prohibit these kinds of observables and
dynamics.

\item Recently there is a proof that QAC is equivalent to the more
standard quantum computation~\cite{Aharonov04}. That this
equivalence generates no contradiction between the
hypercomputability of QAC and the Turing computability of standard
quantum computation can be seen from the facts that (i) such a proof
of equivalence is only valid for finite Hilbert spaces; and (ii) the
hypercomputability of QAC is also based on its probabilistically
correct nature which escapes the jurisdiction of Cantor's diagonal
arguments as discussed in Subsection~\ref{outside}

\item  In the early days there was a claim by
Tsirelson that our proposed algorithm would not
work~\cite{Tsirelson}. We have immediately
replied~\cite{kieuReplyTsirelson} and pointed out the fallacy of the
criticism (but somehow the criticism is still propagating in some
circle, where our reply is not acknowledged). It suffices to point
out here that were Tsirelson correct then the quantum adiabatic
theorem itself would have been wrong, not just our algorithm. But,
of course, there is no basis to suspect the mathematically and
physically proven QAT.

\item Our algorithm requires the ability to generate/simulate
polynomials in the destruction and creation operators $a$ and
$a\dagger$~(\ref{SHOoperators}) to arbitrarily finite orders. These
can be simulated in quantum optics~\cite{Lloyd} with just mirrors,
beam splitters, phase shifts, squeezing and Kerr non-linearity --
despite the appearance of higher powers of the linear momentum
operators~\footnote{In the simple harmonic oscillators, $p \approx
i(a -a^\dagger)$.} which has led some people to wrongly think that
such polynomials could not be generated physically.

\item Others cling to the fact that Grover's quantum search~\cite{Grover}
of an unstructured database of size $M$ would need a number of
computation steps of order $O(\sqrt{M})$ to claim that our algorithm
would then need an infinite time to search the infinite underlying
space. It suffices to note that our search is based on different
principles than those of Grover's. In fact, there do exist in the
literature some other quantum search algorithms, for
example~\cite{othersearch}, that are different to ours and Grover's
and yet whose complexity are not bounded by the size of the search
database as O($\sqrt{M}$).

\item Yet others have erroneously based on a flawed application and
interpretation of the energy-time uncertainty principle to claim
that we could not measure the precise energy of the ground state of
our algorithm in a finite time.
\begin{itemize}
\item This is a flawed application of the uncertainty principle because
our algorithm does not require measurement of {\em arbitrarily}
small energy differences. As a matter of facts, the final energy
spectrum of $H_P$ is {\em integer-valued}, and as such one only
needs to distinguish energies that are separated by at least one
integer unit, not by an arbitrarily small amount. Such measurement
is allowed by the uncertainty principle and can be completed in a
finite time when the energy uncertainties are sufficiently smaller
than the energy unit by which the energy eigenstates are at least
separated.

\item That this is a flawed interpretation of the uncertainty principle
can be seen from the results~\cite{Aharonov61, Aharonov02} that the
energy can be measured in an arbitrarily short time for our
algorithm through the instantaneous measurements of the (Fock)
occupation numbers, one for each variables in the original
Diophantine equation, of the final ground state. These measurements
are compatible with the energy measurement because the final
Hamiltonian $H_P$ commutes, by construction, with the Fock
occupation number operators. And once obtained, these integer-valued
occupation numbers can be simply substituted into the original
Diophantine to reveal the lack or existence of integer solution for
the equation, and also to evaluate the energy of the ground state.
Thus the ground state energy need not be measured directly, even
though it can be done with some care taken to calibrate the energy
zero point.
\end{itemize}

\item This last point leads us to the question whether such
integer-valued occupation numbers can be obtained for very large
values if the universe is finite. The short answer to this question
is obviously NO if the universe is in fact finite,
see~\cite{kieuReplySrikanth} in replying to~\cite{Srikanth}. But
this could be a misleading question to be asked.

Is the universe finite? We do not know for sure, but this is an
important physics question and will surely be investigated and
debated thoroughly in the years to come. But if it is then all
computation, including Turing computation, must be limited, let
alone any hypercomputation. Full stop. With finite physical
resources, any Turing machine can physically compute only some
finite number of the binary digits of any real number. For any
number larger than this physical limit, only abstract mathematical
representations can exist. In this way, we would have to conclude
that $\pi$, for example, were Turing {\em noncomputable} too! Also,
`most' rational numbers would have been classified noncomputable!
Clearly, this is too restrictive and not very useful a discussion of
computable numbers. In fact, with such restriction, one would not
need the concepts of effective computation and of recursive
functions in general. Nor would one need the thesis of Church Turing
at all -- let alone hoping that the physical finiteness of the
universe would support the thesis itself.

In short, the physical finiteness of the universe should not impose
any limitations on hypercomputation more than those which it would
already impose on Turing computation since, in the end, {\em all}
computation is physical. Because of this indiscrimination, it is
logically inconsistent and simply wrong to use the finiteness
arguments to rule out hypercomputation while still maintaining and
defending the validity of Turing computation and the Church-Turing
thesis. On the other hand, the physical finiteness should not and
cannot stop us from investigating hypercomputation as it has not
deterred us from studying Turing computation (or mathematics in
general).

\item The physical finiteness of the universe would of course impose
some upper limit on the number of dimensions one can physically
realise. But as we know when in a Turing computation the end of a
Turing tape has been reached and cannot be lengthened further due to
lack of resources, we should also know when the upper dimensions of
the computation Hilbert space have been physically arrived at. At
that point, the computation would have to be abandoned before we
could obtain the final result. At no time, however, the physical
finiteness of the universe should lead us to the wrong computation
result; it simply would not allow us to complete the computation for
some group of Diophantine equations.

All this depends on whether we could find a probability measure to
quantify, for our final results, the uncertainty associated with a
finite number of dimensions of the underlying Hilbert space. We hope
to pursue this issue elsewhere.

\item What if quantum mechanics is wrong?  A successful theory like
quantum mechanics should not be considered wrong, it could only have
a limited domain of validity, outside which another physical theory
should take over. The question then is whether that validity domain
is good enough for our proposed algorithm. The answer to this
question is unknown at present and as long as quantum mechanics
still proven to be consistently applicable to all phenomena, without
any exception, that can be observed -- as is the situation
presently. The alternative to an exploration of a theory, including
quantum mechanics, for fear that the theory might not be applicable
is to do nothing meaningful.
\end{enumerate}

\subsection{Possible difficulties faced in a physical implementation}
We could perhaps implement the algorithm, among other methods, with
quantum optical apparatuses in which a beam of quantum light is the
physical system on which final measurements are performed and the
number of photons is the quantity measured. The Hamiltonians could
then be physically simulated by various components of mirrors, beam
splitters, Kerr-nonlinear media (with appropriate efficiency),
squeezing~\cite{Lloyd, kieu-intjtheo, kieu-contphys}. We should
differentiate the relative concepts of energy involved in this case;
a final beam state having one single photon, say, could correspond
to a {\em higher} energy eigenstate of $H_P$ than that of a state
having more photons! Only in the final act of measuring photon
numbers, the more-photon state would transfer more energy in the
measuring device than the less-photon state.

The road leading to such a realisation, even if it is not explicitly
prohibited by some known or to-be-found physical principles, is of
course not without problems.
\begin{enumerate}
\item A possible problem is that the Hamiltonians which we need to be
simulated in the optical apparatuses are only {\em effective
Hamiltonians} in that their descriptions are only valid for certain
range of number of photons. When there are too many photons, a
mirror, for example, may respond in a different way from when only a
few photons impinge on it, or the mirror may simply melt down. That
is, other more fundamental processes/Hamiltonians different than the
desirable effective Hamiltonians would take over beyond certain
limit in the photon numbers.

This situation is not unlike that of the required unboundedness of
the Turing tape. In practice, we can only have a finite Turing
tape/memory/register; and when the register is overflowed we would
need to extend it. Similar to this, we would have to be content with
a finite range of applicability for our simulated Hamiltonians. But
we should also know the limitation of this applicability range and
be able to tell when in a quantum computation an overflow has
occurred -- that is, when the range of validity is breached. We
could then use new materials with extended range of (photon number)
applicability.

On the other hand, however, one should note that infinite dimensions
are common and essential in Quantum Mechanics. For example, it is
well known that no finite-dimensional matrices can possibly satisfy
the commutator~\footnote{Were $x$ and $p$ finite matrices, the trace
of the lhs will vanish (as ${\rm tr}(x\,p) = {\rm tr}(p\,x)$) while
the trace of the identity matrix on the rhs does not!}
\[ x\,p - p\,x = i\hbar\,{\bf 1}.\]

\item The ability to control or suppress environmental effects is
also another crucial requirement for the implementation of our
quantum algorithm. The coupling with the environment causes some
fluctuations in the energy levels. Ideally, the size of these
fluctuations should be controlled and reduced to a degree that is
smaller than the size of the smallest energy gap such as not to
cause transitions out of the adiabatic process. This may present a
difficulty for our algorithm, but maybe one of technical nature
rather than of principle. Even though these fluctuations in the
energy levels are ever present and cannot be reduced to zero, there
is no physical principle, and hence no physical reason, why their
sizes cannot be reduced to a size smaller than some required scale.

More specifically are fluctuations in the values of the (integer)
coefficients of the Diophantine equations being implemented in the
Hamiltonian $H_P$~\footnote{This point has been raised on separate
occasions by Martin Davis (2003), Stephen van Enk (2004) and Andrew
Hodges (2004), see also~\cite{Davis04}.}. Those fluctuations will
result in a wrong representation of the Diophantine equation and
thus in erroneous outcomes. If the fluctuations are systematic then
the Diophantine equation cannot be represented correctly at all, and
we then cannot do anything much. The case we wish to investigate
elsewhere is the {\em stochastic} but non-biased fluctuations, so
that the time averages of the coefficients in $H_P$ have the correct
values. Of particular interest is how the ground state and its
measurement probability would be affected by and would scale with
the sizes of such fluctuations in the coefficients, which could be
reduced arbitrarily even though non-vanishing.

\item Recently we have found an interesting and
intimate connection between our algorithm and the so called Mott
insulator - superfluid quantum phase transitions, which have
recently been experimentally confirmed~\cite{Greiner02}. Quantum
phase transitions~\cite{Sachdev} are purely due to quantum
fluctuations, unlike other types of phase transitions, such as
liquid-vapour transitions, which are due entirely to the thermal
fluctuations. In principle, quantum fluctuations can thus take place
even at zero absolute temperature. This connection thus irrefutably
demonstrates the possibility of physical implementations for our
algorithm, at least for certain classes of Diophantine equations.
The details of the connection will be published elsewhere. It
suffices to mention here that such a connection is possible because
the underlying process for quantum phase transitions is also the
quantum adiabatic process that governs the dynamics of the quantum
adiabatic algorithm.
\end{enumerate}

\section{Concluding remarks}
In this paper we have given an overview of the working of an
adiabatic quantum algorithm for Hilbert's tenth problem. Numerical
simulations for some simple Diophantine equations have also been
reported together with an explanation how we should cope with the
required dimensionally-unbounded Hilbert space.

Our algorithm is not just another quantum algorithm in the sense
that it can substantially speed up what can be computed on classical
computers, but it presents a new paradigm of computation. It is
arguably in the area of hypercomputation, the kind of computability
beyond that delimited by the Church-Turing thesis. This is due to
the facts that: (i) it belongs to the class of probabilistically
correct algorithms which are not subject to Cantor's diagonal
arguments as we have pointed out above in Subsection~\ref{outside};
and (ii) it is based on the most general principles of quantum
mechanics which are definitely and irrefutably non-recursive as can
be seen through the class of Schr\"odinger
equations~(\ref{Schrodinger}) and through their ability to generate
truly random numbers.

The fact that our algorithm is ``only" probabilistically correct can
be understood as a necessity and a consistency condition when the
outcomes of such an algorithm cannot, in principle, be verified by
any other means. The algorithm gives the $k$-tuple at which the
square of a Diophantine polynomial assumes it smallest value. While
the existence of a solution can be verified by a simple
substitution, the indication of no solution cannot be verified by
any other finite recursive means at all -- thus the need of some
probability measure to quantify the accuracy of the derived
conclusion. However, it is important and useful that this
probability is not only known but can also be predetermined with an
arbitrary value in advance.

The algorithm is based on the quantum adiabatic theorem, which
asserts that a particular eigenstate of a final-time Hamiltonian,
even in dimensionally {\em infinite} spaces, could be found
mathematically and/or physically {\em in a finite time}. This is a
remarkable property, which is enabled by quantum interference and
quantum tunnelling with complex-valued probability amplitudes, and
allows us to find a needle in an infinite haystack, in principle!
Such a property is clearly {\em not} available for recursive search
in an unstructured infinite space, which, in general, cannot
possibly be completed in a finite time, in contrast to the quantum
scenario.

This is what we mean by ``an infinite search in a finite time": {\em
Of course, the spread of a quantum wavefunction in an adiabatic
process for a finite time can only cover essentially a finite domain
of the underlying infinite space, but the finite domain so covered
is the relevant domain, as guaranteed by the quantum adiabatic
theorem, for a successful location of the global minimum (which is
the search objective) in a finite time -- whence the locating result
is necessarily subject to some probability of error which can
nevertheless be arbitrarily reduced.}

That is the point. Just like water in a porous landscape always
accumulates at the lowest point after some finite time even when the
size of the landscape is infinite (since the water need not explore
the whole space), a quantum state (starting from appropriate initial
conditions) could also accumulate (adiabatically and
probabilistically) at the global minimum (of some bounded-from-below
Hamiltonian) -- even in an infinite Hilbert space -- given
sufficient but finite time for it to perform its quantum tunneling
instinct.

The failure of Cantor's diagonal arguments to rule out the use of
probabilistic procedure for hypercomputation leaves open the
possibility that we could employ a single and universal, albeit
probabilistically correct, algorithm to compute some recursively
non-computable, {\em provided we are prepared to be wrong -- with an
error probability that can be made arbitrarily small}. This fact was
recognised in one form or another by G\"odel and Turing themselves a
long time ago (see~\cite{kieu05} for some discussions and
quotations). And we would like to think that that probabilistic
hypercomputation possibility is the quantum adiabatic algorithm we
have proposed for Hilbert's tenth problem and its associated class
of problems. Perhaps this hypercomputability of quantum adiabatic
processes could be generalised further to a wider class of finitely
refutable problems?

Elsewhere, in the earlier stage of development of the algorithm, I
thought and accordingly stated that the ability to implement
dimensionally infinite Hamiltonians was absolutely
necessary~\cite{kieu-mindmach}. While that ability is sufficient and
certainly of great help for our hypercomputational algorithm and
others~\cite{Ziegler04}, I am now of the opinion that it would not
be quite absolutely necessary if we could quantify the uncertainties
associated with truncated Hilbert spaces by some error probabilities
for the outcomes. We will pursue this issue elsewhere.

Intimately linked with this computability study are the physical
phenomena of quantum phase transitions (QPT). The fact that certain
QPT is the physical realisation of instances of our algorithms for
Hilbert's tenth problem is surprising, extraordinary and of great
consequence. It does not only illustrate that certain instances of
the algorithm are physically implementeble but also raises the
prospect of a much deeper connection between QAC, quantum
computability and QPT.


These studies into the limit of mathematics and generalised
noncomputability and undecidability set the bounds for computation
carried out by mechanical (including quantum mechanical) processes,
and in so doing help us to understand much better what can be so
computed. It should enable us to algorithmically resolve many
important and interesting mathematical problems. It would further
contribute to other fields of philosophy and artificial intelligence
-- in the debate, for example, whether human minds can be simulated
(or even be replaced) by machines and computers.

\section*{Notes added}
During the course of writing this paper I have learned about Martin
Davis' comments~\cite{DavisThisIssue} about my not responding to
Andrew Hodges' critique of the algorithm as posted on FOM (a forum
on Foundation of Mathematics moderated by Davis). In fact, we have
had not only with Hodges but also with other people quite a few
discussions, for which I am very grateful. However, after some
initial exchanges, I have explicitly stated on FOM to the effect
that I thought that the forum, while useful for clearing up some
misunderstandings, was neither the appropriate medium nor a primary
publication place for settling deep disagreements. I have also
proposed to these people that their opposing arguments should be
published for the record. (And this should also apply to ``the
leading experts in quantum computation" who have advised Davis
against my proposed algorithm, as I have never seen their arguments
in print -- except those in~\cite{Tsirelson},~\cite{Srikanth},
respectively to which I have already replied
in~\cite{kieuReplyTsirelson},~\cite{kieuReplySrikanth}.) I myself
have, for the record, addressed above in this paper and also
elsewhere~\cite{kieu05} some of the points (that I know of) raised
by Hodges and others about the algorithm.

\section*{Acknowledgments} I am indebted to Alan Head, Peter
Hannaford, Toby Ord and Andrew Rawlinson for discussions and
continuing support. I would also like to acknowledge helpful
discussions with Enrico Deotto, Ed Farhi, Jeff Goldstone and Sam
Gutmann during a visit to MIT in 2002, and with Peter Drummond and
Peter Deuar; these discussions have helped clarifying the issues
discussed in this paper.

\bibliography{adiabatic,ait}

\begin{thebibliography}{10}

\bibitem{Aharonov04}
D.~Aharonov, W.~van Dam, J.~Kempe, Z.~Landau, S.~Lloyd, and O.~Regev.
\newblock Adiabatic computation is equivalent to standard quantum computation.
\newblock {\tt ArXiv:quant-ph/0405098}, 2004.

\bibitem{Aharonov61}
Y.~Aharonov and D.~Bohm.
\newblock Time in the quantum theory and the uncertainty relation for time and
  energy.
\newblock {\em Phys. Rev.}, 122:1649--1658, 1961.

\bibitem{Aharonov02}
Y.~Aharonov, S.~Massar, and S.~Popescu.
\newblock Measuring energies, estimating {H}amiltonians, and the time-energy
  uncertainty relation.
\newblock {\em Phys. Rev. A}, 66:052107, 2002.

\bibitem{bernsteinetal}
E.~Bernstein and U.~Vazirani.
\newblock Quantum complexity theory.
\newblock {\em SIAM J. Comp.}, 26:1411, 1997.

\bibitem{CaludeBookRandom}
C.S. Calude.
\newblock {\em Information and Randomness: An Algorithmic Perspective}.
\newblock Springer-Verlag, Berlin Heidelberg, 2nd edition, 2002.

\bibitem{CaludePavlov}
C.S. Calude and B.~Pavlov.
\newblock Coins, quantum measurements, and turing's barrier.
\newblock {\em Quantum Information Processing}, 1:107--127, 2002.

\bibitem{mountaintop}
J.L. Casti.
\newblock {\em Mathematical Mountaintops: The Five Most Famous Problems of All
  Time}.
\newblock Oxford University Press, New York, 2001.

\bibitem{CastiDePauli}
J.L. Casti and W.~DePauli.
\newblock {\em G\"odel: A Life of Logic}.
\newblock Perseus Publishing, Cambridge, Massachusetts, 2000.

\bibitem{Chaitin:1987}
G.J. Chaitin.
\newblock {\em Algorithmic Information Theory}.
\newblock Cambridge University Press, Cambridge, 1987.

\bibitem{ChaitinBook05}
G.J. Chaitin.
\newblock {\em Meta Math! The Quest for $\Omega$}.
\newblock Pantheon Books, New York, 2005.

\bibitem{farhi2}
A.M. Childs, E.~Farhi, and J.~Preskill.
\newblock Robustness of adiabatic quantum computation.
\newblock {\em Phys. Rev. A}, 65:012322, 2002.

\bibitem{Copeland}
J.~Copeland.
\newblock Hypercomputation.
\newblock {\em Minds and Machines}, 12:461--502, 2002.

\bibitem{Cotogno}
P.~Cotogno.
\newblock Hypercomputation and the physical {C}hurch-{T}uring thesis.
\newblock {\em Brit. J. Phil. Sci.}, 54:181--223, 2003.

\bibitem{othersearch}
S.~Das, R.~Kobes, and G.~Kunstatter.
\newblock Energy and efficiency of adiabatic quantum search algorithms.
\newblock {\em J. Phys. A}, 36:2839--46, 2003.

\bibitem{Davis02}
M.~Davis.
\newblock {\em Engines of Logic: Mathematicians and the Origin of the
  Computer}.
\newblock W.W. Norton, New York, 2001.

\bibitem{Davis04}
M.~Davis.
\newblock The myth of hypercomputation.
\newblock In C.~Teuscher, editor, {\em Alan Turing: Life and Legacy of a Great
  Thinker}, pages 195--212. Springer, Berlin, 2004.

\bibitem{DavisThisIssue}
M.~Davis.
\newblock Why there is no such discipline as hypercomputation.
\newblock This issue, 2005.

\bibitem{shannon}
K.~de~Leeuw, E.F. Moore, C.E. Shannon, and N.~Shapiro.
\newblock Number~34 in Automata Studies Annals of Mathematics Studies.
  Princeton University Press, Princeton, 1956.

\bibitem{qac}
E.~Farhi, J.~Goldstone, S.~Gutmann, and M.~Sipser.
\newblock Quantum computation by adiabatic evolution.
\newblock {\tt ArXiv:quant-ph/0001106}, 2000.

\bibitem{Greiner02}
M.~Greiner, O.~Mandel, T.W. H\"ansch, and I.~Bloch.
\newblock Collapse and revival of the matter wave field of a {B}ose-{E}instein
  condensate.
\newblock {\em Nature}, 419:51--54, 2002.

\bibitem{Grover}
L.K. Grover.
\newblock Quantum mechanics helps in searching for a needle in a haystack.
\newblock {\em Phys. Rev. Lett.}, 79:325--328, 1997.

\bibitem{kieuReplyTsirelson}
T.D. Kieu.
\newblock Reply to `{T}he quantum algorithm of {K}ieu does not solve the
  {H}ilbert's tenth problem'.
\newblock {\tt ArXiv:quant-ph/0111020}, 2001.

\bibitem{kieu-mindmach}
T.D. Kieu.
\newblock Quantum hypercomputation.
\newblock {\em Minds and Machines}, 12:541--561, 2002.

\bibitem{kieu-contphys}
T.D. Kieu.
\newblock Computing the non-computable.
\newblock {\em Contemporary Physics}, 44:51--77, 2003.

\bibitem{kieu-spie}
T.D. Kieu.
\newblock Numerical simulations of a quantum algorithm for {H}ilbert's tenth
  problem.
\newblock In Eric Donkor, Andrew~R. Pirich, and Howard~E. Brandt, editors, {\em
  Proceedings of SPIE Vol. 5105 {\it Quantum Information and Computation}},
  pages 89--95. SPIE, Bellingham, WA, 2003.

\bibitem{kieuFull}
T.D. Kieu.
\newblock Quantum adiabatic algorithm for {H}ilbert's tenth problem: I. {T}he
  algorithm.
\newblock \texttt{ArXiv:quant-ph/0310052}, 2003.

\bibitem{kieu-intjtheo}
T.D. Kieu.
\newblock Quantum algorithms for {H}ilbert's tenth problem.
\newblock {\em Int. J. Theor. Phys.}, 42:1451--1468, 2003.

\bibitem{kieuReplySrikanth}
T.D. Kieu.
\newblock Finiteness of the universe and computation beyond {T}uring
  computability.
\newblock {\tt ArXiv:quant-ph/0403045}, 2004.

\bibitem{kieu-royal}
T.D. Kieu.
\newblock A reformulation of {H}ilbert's tenth problem through quantum
  mechanics.
\newblock {\em Proc. Roy. Soc.}, A 460:1535--1545, 2004.

\bibitem{kieu05}
T.D. Kieu.
\newblock An anatomy of a quantum adiabatic algorithm that transcends the
  {T}uring computability.
\newblock {\em Int. J. Quantum Info.}, 3:177--182, 2005.

\bibitem{Lloyd}
S.~Lloyd and S.~Braunstein.
\newblock Quantum computation over continuous variables.
\newblock {\em Phys. Rev. Lett.}, 82:1784, 1999.

\bibitem{hilbert10}
Y.V. Matiyasevich.
\newblock {\em Hilbert's Tenth Problem}.
\newblock MIT Press, Cambridge, Massachussetts, 1993.

\bibitem{Matiyasevich04}
Y.V. Matiyasevich.
\newblock Diophantine flavour of {K}olmogorov complexity.
\newblock {\tt http://at.yorku.ca/cgi-bin/amca/cani-11}, 2004.

\bibitem{messiah}
A.~Messiah.
\newblock {\em Quantum Mechanics}.
\newblock Dover, New York, 1999.

\bibitem{nielsen}
M.A. Nielsen.
\newblock Computable functions, quantum measurements, and quantum dynamics.
\newblock {\em Phys. Rev. Lett.}, 79:2915--8, 1997.

\bibitem{qcbook}
M.A. Nielsen and I.L. Chuang.
\newblock {\em Quantum Computation and Quantum Information}.
\newblock Cambridge University Press, Cambridge, 2000.

\bibitem{Toby02}
T.~Ord.
\newblock Hypercomputation: Computing more than the {Turing} machine.
\newblock Honours Thesis, University of Melbourne, Melbourne, Australia,
  September 2002. Also available at {\tt ArXiv:math.LO/0209332}, 2002.

\bibitem{Toby05}
T.~Ord.
\newblock Hypercomputation: computing more than the {T}uring {M}achine.
\newblock This issue, 2005.

\bibitem{ordkieu-omega}
T.~Ord and T.D. Kieu.
\newblock On the existence of a new family of {Diophantine} equations for
  {$\Omega$}.
\newblock {\em Fundmenta Informaticae}, 56:273--84, 2003.

\bibitem{OrdKieu-qubit}
T.~Ord and T.D. Kieu.
\newblock Using biased coins as oracles.
\newblock {\tt Arxiv:cs.OH/0401019}, 2004.

\bibitem{OrdKieu-diag}
T.~Ord and T.D. Kieu.
\newblock The diagonal method and hypercomputation.
\newblock {\em Brit. J. Phil. Sci.}, 56:147--156, 2005.

\bibitem{PourEl}
M.B. Pour-El and J.I. Richards.
\newblock {\em Computability in Analysis and Physics}.
\newblock Sringer-Verlag, Berlin-Heidelberg, 1989.

\bibitem{ReedSimon}
M.~Reed and B.~Simon.
\newblock {\em Methods of Modern Mathematical Physics: IV Analysis of
  Operators}.
\newblock Academic Press, San Diego, 1978.

\bibitem{probtheory}
Y.A. Rozanov.
\newblock {\em Probability Theory: A Concise Course}.
\newblock Dover, New York, 1977.

\bibitem{Sachdev}
S.~Sachdev.
\newblock {\em Quantum Phase Transitions}.
\newblock Cambridge University Press, Cambridge, 1999.

\bibitem{Santos:1971}
E.S. Santos.
\newblock Computability by probabilistic {T}uring machines.
\newblock {\em Trans. Am. Math. Soc.}, 159:165--184, 1971.

\bibitem{columbia2}
A.~Sicard, J.~Ospina, and M.~V\'elez.
\newblock Numerical simulations of a possible hypercomputational quantum
  algorithm.
\newblock In B.~Ribeiro et~al., editors, {\em Proceedings of the International
  Conference on Adaptive and Natural Computing Algorithms}. Springer Wien
  NewYork, 2005.

\bibitem{columbia}
A.~Sicard, M.~V\'elez, and J.~Ospina.
\newblock Hypercomputation based on quantum computing.
\newblock {\tt ArXiv:quant-ph/0406137}, 2004.

\bibitem{Srikanth}
R.~Srikanth.
\newblock Computable functions, the {C}hurch-{T}uring {T}hesis and the quantum
  measurement problem.
\newblock {\tt ArXiv:quant-ph/0402128}, 2004.

\bibitem{Srinivasan04}
R.~Srinivasan and H.P. Raghunandan.
\newblock On the existence of truly autonomic computing systems and the link
  with quantum computing.
\newblock {\tt ArXiv:cs.LO/0411094}, 2004.

\bibitem{Stannett}
M.~Stannett.
\newblock Hypercomputation is experimentally irrefutable.
\newblock Sheffield University Department of Computer Science Report TR
  CS-01-04, 2001.

\bibitem{Tsirelson}
B.~Tsirelson.
\newblock The quantum algorithm of {K}ieu does not solve the {H}ilbert's tenth
  problem.
\newblock {\tt ArXiv:quant-ph/0111009}, 2001.

\bibitem{Welch04}
P.D. Welch.
\newblock On the possibility, or otherwise, of hypercomputation.
\newblock {\em Brit. J. Phil. Sci.}, 55:739--746, 2004.

\bibitem{Ziegler04}
M.~Ziegler.
\newblock Does quantum mechanics allow for infinite parallelism?
\newblock {\tt ArXiv:quant-ph/0410141}, 2004.

\end{thebibliography}
\bibliographystyle{plain}

\end{document}